\documentclass[aps,prd,amsmath,amssymb,showpacs]{revtex4}
\usepackage{graphicx}
\usepackage{dcolumn}
\usepackage{morefloats}
\usepackage{color}
\usepackage{slashed}
\usepackage{bbm}

\begin{document}
\title{Towards the establishment of the light $J^{P(C)}=1^{-(+)}$ hybrid nonet}

\author{Lin Qiu$^{1,2}$\footnote{{\it E-mail address:} qiulin@ihep.ac.cn}, 
and Qiang Zhao$^{1,2,3}$\footnote{{\it E-mail address:} zhaoq@ihep.ac.cn}}

\affiliation{$^1$ Institute of High Energy Physics and Theoretical Physics Center
for Science Facilities,\\
         Chinese Academy of Sciences, Beijing 100049, China}

\affiliation{$^2$ School of Physical Sciences, University of Chinese Academy of
Sciences, Beijing 100049, China}

\affiliation{$^3$ China Center of Advanced Science and Technology, Chinese Academy of
Sciences, Beijing 100080, China}

\begin{abstract}

The observation of the light hybrid candidate $\eta_1(1855)$ by the BESIII Collaboration brings great opportunities for advancing our knowledges about exotic hadrons in the light flavor sector. We show that this observation provides a crucial clue for establishing the $J^{P(C)}=1^{-(+)}$ hybrid nonet. Based on the flux tube model picture, the production and decay mechanisms for the $J^{P(C)}=1^{-(+)}$ hybrid nonet in the $J/\psi$ radiative decays into two pseudoscalar mesons are investigated. In the $I=0$ sector, we find that the SU(3) flavor octet and singlet mixing is non-negligible and apparently deviates from the flavor ideal mixing. Since only signals for one isoscalar $\eta_1(1855)$ are observed in the $\eta\eta'$ channel, we investigate two schemes of the nonet structure in which $\eta_1(1855)$ can be either the higher or lower mass state that strongly couples to $\eta\eta'$. Possible channels for detecting the multiplets are suggested. In particular, a combined analysis of the hybrid production in $J/\psi\to VH$, where $V$ and $H$ stand for the light vector mesons and $1^{-(+)}$ hybrid states, may provide further evidence for this nonet structure and finally establish  these mysterious exotic species in experiment.

\end{abstract}

\maketitle
\section{Introduction}

In the conventional quark model mesons are made of quark-anti-quark ($q\bar{q}$) and baryons are made of three quarks ($qqq$). Such a simple picture have made great successes in the description of hadron spectra based on the constituent quark degrees of freedom. Meanwhile, QCD as the fundamental theory for strong interactions predicts the existence of hadrons with more sophisticated structures, namely, exotic hadrons. These states, of which the structures are beyond the conventional quark model, have been a crucial probe for the non-perturbative phenomena of QCD. Among all the exotic candidates, hadrons with such quantum numbers that cannot be accommodated by the conventional quark model, would serve as a ``smoking gun" for the existence of exotic hadrons. In particular, ``hybrid", which contains the explicit excitations of the constituent-like gluonic degrees of freedom, can access the exotic quantum numbers of $J^{PC}=1^{-+}$ as the lowest eigenstates. Its study has always attracted a lot of attention from both experiment and theory.

In Refs.~\cite{besiii-hybrid,{besiii-hybrid-pwa}} the BESIII Collaboration reports the first observation of the $1^{-+}$ isoscalar hybrid candidate $\eta_1(1855)$ in the partial wave analysis of $J/\psi\to \gamma \eta_1(1855)\to\gamma\eta\eta'$. Its mass and width are $(1855\pm 9^{+6}_{-1})$ MeV and $(188\pm 18^{+3}_{-8})$ MeV, respectively. This progress may provide a great opportunity for a better understanding of these mysterious species of the QCD-predicted exotic states.

Historically, evidences for the $1^{-+}$ hybrid were found by various experiments~\cite{SLACHybridFacilityPhoton:1991oug,VES:1993scg,Lee:1994bh,Aoyagi:1993kn,E852:1997gvf,E852:1999xev}, and two light hybrid candidates, $\pi_1(1400)$ and $\pi(1600)$, were reported. However, due to the limited statistics, their existences were far from broadly accepted. A comprehensive review of the early experimental results can be found in Refs.~\cite{Klempt:2007cp,Meyer:2015eta}. Strong indication of  the $1^{-+}$ hybrid $\pi_1(1600)$ is from the COMPASS Collaboration based on their partial wave analysis (PWA) result for $\pi^- p\to p \pi^+\pi^-\pi^-$~\cite{COMPASS:2009xrl,E852:2001ikk,COMPASS:2014vkj,COMPASS:2018uzl}. In a recent detailed analysis~\cite{COMPASS:2018uzl} by COMPASS, it shows that the $\pi_1(1600)$ signal cannot be accounted for by the Deck effect~\cite{Deck:1964hm}. A reanalysis of the COMPASS data with the coupled-channel approach also supports that the $\pi_1(1600)$ signal should be originated from a pole structure in the scattering amplitude~\cite{JPAC:2018zyd}. These results have provided strong evidences for $\pi_1(1600)$ as a well-established $1^{-+}$ hybrid candidate. In contrast, the signals for $\pi_1(1400)$ turn out to be vague. According to the analysis of Ref.~\cite{JPAC:2018zyd}, there is no need for the $\pi_1(1400)$ to be present in the $\rho\pi$ channel.

Phenomenological studies of the $1^{-+}$ hybrid state can be found in the literature. By treating the gluonic excitation as an explicit constituent degree of freedom, phenomenological models were constructed to understand the exotic hadron spectrum or describe the mechanisms for their productions and decays~\cite{Horn:1977rq,Barnes:1982zs}. Among all these efforts, the flux tube model has made a great success in accommodating the broadly adopted quark pair creation (QPC) model for the strong decays of conventional hadrons and the gluonic excitations of QCD exotics~\cite{Isgur:1984bm,Barnes:1995hc,Page:1998gz,Close:2003af}. 
Calculations in the framework of QCD sum rules also provide interesting results on the properties of the light hybrid $\pi_1$ state~\cite{Zhu:1999wg,Huang:2010dc}. In Refs.~\cite{Chen:2010ic} the decay properties are studied for $\pi_1$ and its non-strange isoscalar partner. In Ref.~\cite{Zhang:2019ykd}, it is investigated that an isoscalar with $I^G(J^{PC})=0^+(1^{-+})$ may be formed as a bound state of $\eta \bar{K}K^*$. However, the mass is much lower than $\eta_1(1855)$.

There is no doubt that lattice QCD (LQCD) simulations should play a crucial role in guiding the search for the hybrid states. In Ref.~\cite{Dudek:2013yja} the first systematic LQCD study of the excited isoscalar meson spectra was presented. It is interesting to see the emergence of the mixing patterns between the SU(3) flavor singlet and octet such as the $\eta$ and $\eta'$ mixing. In the $1^{-+}$ hybrid sector, some hints for the mixings between the non-strange and strange configurations are found. Meanwhile, its prediction of the isoscalar $1^{-+}$ hybrid spectrum indicates relatively higher masses than the light axial vector mesons. It implies an unusual behavior of the excitations of the gluonic degrees of freedom in comparison with the orbital excitations within conventional $q\bar{q}$ systems. 

In light of the discovery of $\eta_1(1855)$ by BESIII~\cite{besiii-hybrid,{besiii-hybrid-pwa}} and the LQCD simulations~\cite{Dudek:2011bn,Dudek:2013yja}, we propose a nonet scheme for the $1^{-(+)}$ hybrid states. In this scheme $\pi_1(1600)$ is the $I=1$ state with the lowest mass, and $\eta_1(1855)$ is identified as one of the $I=0$ multiplets. The strange $I=1/2$ partner is assigned to $K^*(1680)$ which is the only strange vector meson found in the vicinity of $1.6\sim 1.9$ GeV mass region. Although the strange hybrid does not have a fixed charge conjugate parity, hence it cannot be easily distinguished from the conventional $q\bar{q}$ vector meson, there is no strong reason that such an exotic object should not exist. Considering the flavor-blind property of QCD, the strange hybrid of $q\bar{s}\tilde{g}$ should at least share similar dynamics as the $I=1$ partner $\pi_1(1600)$. In Ref.~\cite{Dudek:2013yja} the mass splitting between the  flavor singlet and octet is found to be significant. This is attributed to the important effects from the quark annihilations in the $I=0$ sector.

As follows, we first analyse the mixing between $\eta_1(1855)$ and its isoscalar partner, and the mass relations among the $1^{-(+)}$ hybrid nonet. Two schemes, in which $\eta_1(1855)$ is assigned to be either the higher or lower mass state in the $I=0$ sector, are explored based on the flux tube model picture. Phenomenological consequences will be discussed in their productions and decays in several typical processes. A brief summary will be given in the end.

\section{Productions and decays of the $1^{-(+)}$ hybrid states}

\subsection{Emergence of the $1^{-(+)}$ hybrid nonet}

On the SU(3) flavor basis the light hybrid mesons are described by a pair of $q\bar{q}$ associated by gluonic quasiparticle excitations. Taking the flux tube model picture, the $q\bar{q}$ inside hybrid mesons are separated static color sources and they are connected by the gluonic flux tube to form an overall color singlet. The transverse oscillations of the flux tube that manifests the explicit effective gluonic degrees of freedom, will give rise to energy spectrum of the hybrid mesons. As studied in the literature, the lowest energy flux tube motion has  $J_g^{PC}=1^{+-}$. Namely, the lightest hybrid multiplet can be formed by the relative $S$-wave coupling between a gluonic lump of $J_g^{PC}=1^{+-}$ and a $S$-wave $q\bar{q}$ pair. With the total gluon spin $J_g^{PC}=1^{+-}$, the lowest hybrid multiplets can be obtained: $(0, \ 1, \ 2)^{-+}, \ 1^{--}$~\cite{Bali:2003jq,Dudek:2011bn}. Alternatively, in the constituent gluon picture the lowest energy flux tube excitation can be described by the motion of quasigluon in a $P$ wave with respect to the $S$-wave $q\bar{q}$.

The gluonic excitations additive to the $S$-wave constituent $q\bar{q}$ configuration suggests that for each $S$-wave $q\bar{q}$ pair, there should exist an SU(3) flavor nonet as the eigenstates of the corresponding Hamiltonian. For the same coupling mode involving the gluonic lump, these states can be related to each other by the Gell-Mann-Okubo mass relation similar to that for the ground states in the $q\bar{q}$ scenario. This conjecture may have a caveat when the strange multiplets are included. Since the charged and strange states do not have the fixed $C$ parity, it may raise the question whether a nonet scheme makes sense or not. Note that signals for charged $\pi_1(1600)$ have been seen in the decay channels of $\rho^0\pi^-$~\cite{COMPASS:2009xrl} and $\eta'\pi^-$~ \cite{E852:2001ikk,COMPASS:2014vkj}. Similar dynamics should appear in the strange sector and a nonet structure among the $1^{-(+)}$ multiplets should be a good guidance for a better understanding of the underlying dynamics.

Taking the $1^{-+}$ hybrid as an example, it should contain flavor multiplets as follows:
\begin{eqnarray}
\pi_1^+, \ \pi_1^-, \ \pi_1^0& : & u\bar{d}\tilde{g}, \ d\bar{u}\tilde{g}, \ \frac{1}{\sqrt{2}}(u\bar{u}-d\bar{d})\tilde{g} \ ,\\
\eta_{1}^{(8)} & : & \frac{1}{\sqrt{6}}(u\bar{u}+d\bar{d}-2s\bar{s})\tilde{g} \ ,\\
\eta_{1}^{(1)} & : & \frac{1}{\sqrt{3}}(u\bar{u}+d\bar{d}+s\bar{s})\tilde{g} \ ,\\
K^{*+}, \ K^{*0}, \ K^{*-}, \ \bar{K}^{*0} & : &  u\bar{s}\tilde{g}, \ d\bar{s}\tilde{g}, \ s\bar{u}\tilde{g}, \ s\bar{d}\tilde{g} \ ,
\end{eqnarray}
where $\tilde{g}$ represents the gluonic lump with $J_g^{PC}=1^{+-}$. For the flavor octet $\eta_{1}^{(8)}$ and singlet $\eta_{1}^{(1)}$ with isospin $I=0$, they may mix with each other to form the corresponding physical states similar to the familiar $\eta$-$\eta'$ mixing.

Considering the mixing between the hybrid flavor singlet and octet, the physical states can be expressed as
\begin{eqnarray}\label{mixing-eta1}
\left(
\begin{array}{c}
\eta_{1L} \\
\eta_{1H}
\end{array}
\right)&=&
\left(
\begin{array}{cc}
\cos\theta & -\sin\theta \\
\sin\theta & \cos\theta
\end{array}\right)
\left(
\begin{array}{c}
\eta_{1}^{(8)} \\
\eta_{1}^{(1)}
\end{array}
\right)=
\left(
\begin{array}{cc}
\cos\alpha & -\sin\alpha \\
\sin\alpha & \cos\alpha
\end{array}\right)
\left(
\begin{array}{c}
n\bar{n}\tilde{g} \\
s\bar{s}\tilde{g}
\end{array}
\right) \ ,
\end{eqnarray}
where $\theta$ is the mixing angle between the flavor octet and singlet, and $\alpha$ is the mixing angle defined on the flavor basis $n\bar{n}\tilde{g}$ (with $n\bar{n}\equiv (u\bar{u}+d\bar{d})/\sqrt{2}$) and $s\bar{s}\tilde{g}$.

The Gell-Mann-Okubo relation can  provide a constraint on the mixing angle $\theta$ via the following equation: 
\begin{eqnarray}\label{gell-mann-okubo-rela}
\tan\theta&=& \frac{4m_{K^*}-m_{\pi_1}-3m_{\eta_{1L}}}{2\sqrt{2}(m_{\pi_1}-m_{K^*})} \ ,
\end{eqnarray}
where $\eta_{1L}$ is the lower mass state in Eq.~(\ref{mixing-eta1}) and the sign of $\theta$ can be determined here. We note that the same relation is also satisfied for the quadratic masses. The Gell-Mann-Okubo relation also leads to the following mass relation,
\begin{eqnarray}\label{GMO-mass}
(m_{\eta_{1H}}+m_{\eta_{1L}})(4m_{K^*}-m_{\pi_1})-3m_{\eta_{1H}}m_{\eta_{1L}}&=&8m_{K^*}^2-8m_{K^*}m_{\pi_1}+3m_{\pi_1}^2 \ ,
\end{eqnarray}
which is symmetric for $\eta_{1L}$ and $\eta_{1H}$ although it is the lower mass state $\eta_{1L}$ defined in Eq.~(\ref{mixing-eta1}) to appear in Eq.~(\ref{gell-mann-okubo-rela}). With the masses of $\pi_1(1600)$ and $\eta_1(1855)$ as the input for Eqs.~(\ref{gell-mann-okubo-rela}) and (\ref{GMO-mass}), we are still unable to determine these three quantities, i.e. $\theta$, $m_{\eta_{1H}}/m_{\eta_{1L}}$ and $m_{K^*}$. Also, it is unclear whether $\eta_1(1855)$ is the lower or higher mass state in Eq.~(\ref{mixing-eta1}). However, we will show later that the $\eta\eta'$ channel is informative to impose constraint on the determination of the $1^{-(+)}$ nonet.

\subsection{$1^{-(+)}$ nonet decays into pseudoscalar meson pairs}\label{subsect-nonet-decay}

\begin{figure}
\includegraphics[width=2.in]{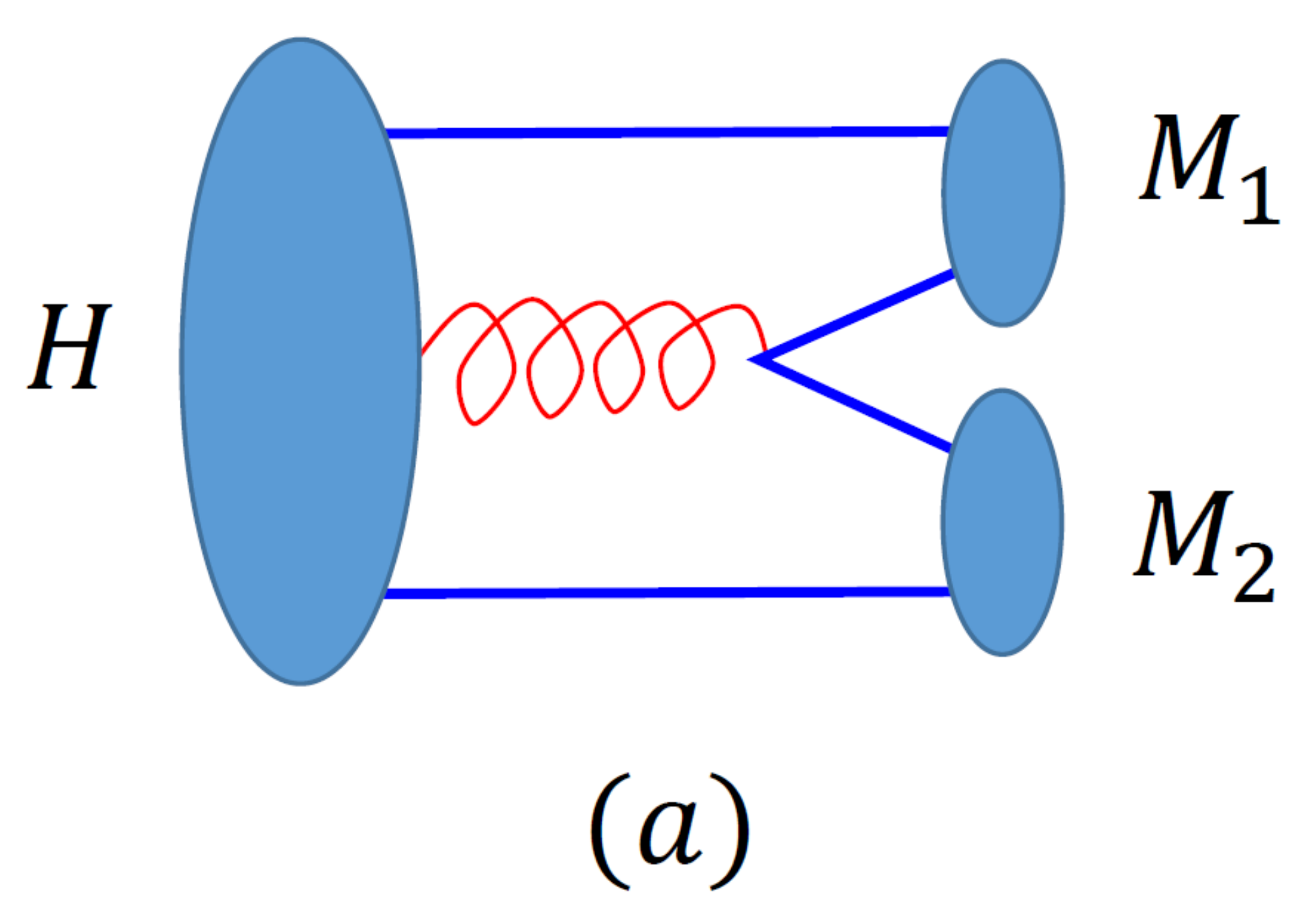}
\includegraphics[width=2.in]{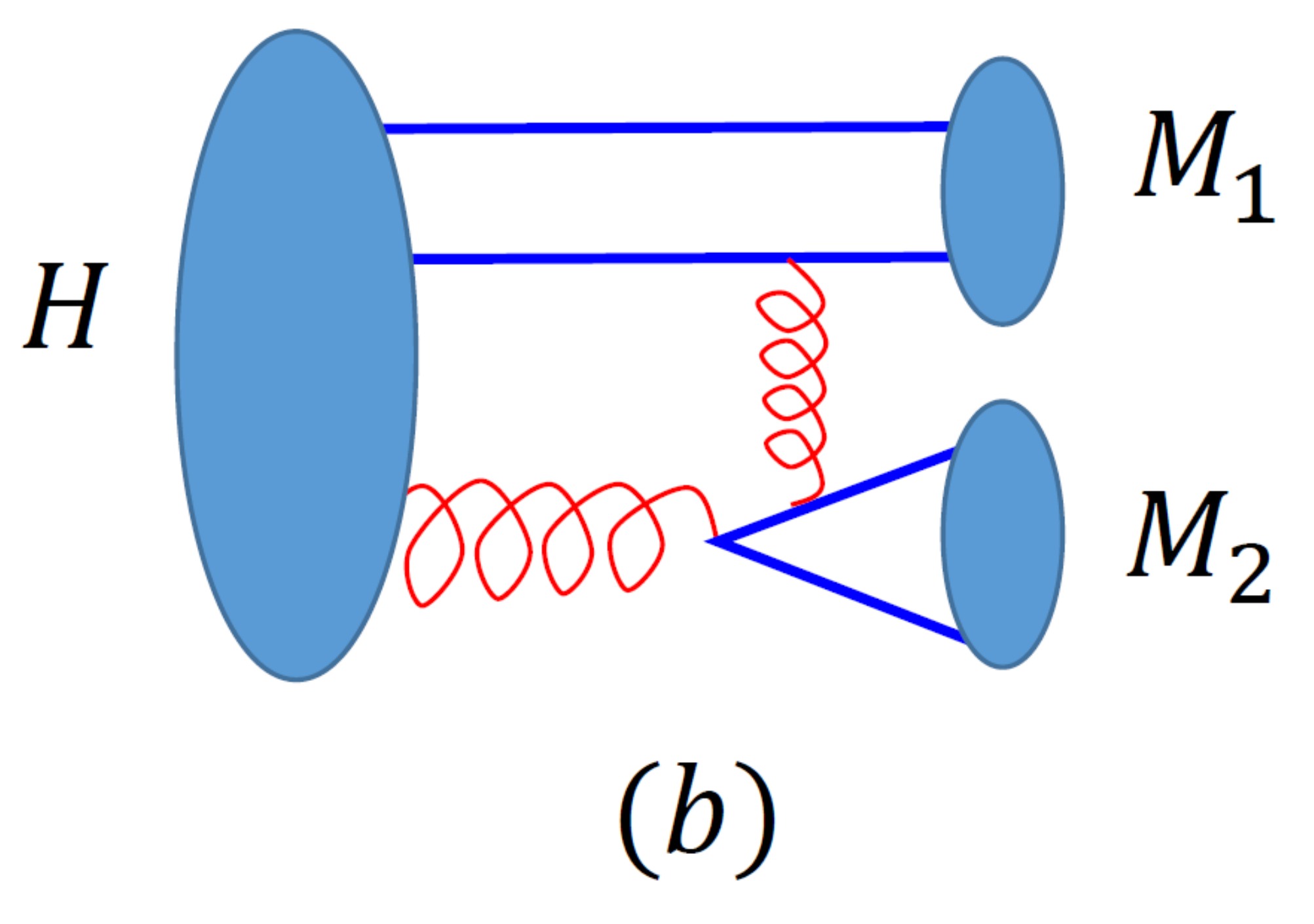}
\caption{Illustration of the $1^{-(+)}$ isoscalar hybrid decays into two mesons.}
\label{fig-1}
\end{figure}

The flavor-blindness of the strong interactions also allows us to relate the SU(3) decay channels together~\cite{Close:2000yk,Close:2005vf,Zhao:2007ze,Zhao:2006cx}. Considering the two-body decay of $\eta_{1L}$ and $\eta_{1H}$ into pseudoscalar meson pair $PP'$~\footnote{Due to Bose symmetry, the $1^{-+}$ hybrid cannot decay into two identical mesons. Namely, the decays into $\pi^0\pi^0$, $\eta\eta$, and $\eta'\eta'$ are forbidden. Also, the $G$-parity conservation will forbid the decays of $\pi_1$ and $\eta_1$ into $\pi\pi$ and $K\bar{K}$. }, two independent transition mechanisms can be identified and they are illustrated by Fig.~\ref{fig-1} (a) and (b). The transition of Fig.~\ref{fig-1} (a) represents the flux tube string breaking with the quark pair creation. It is similar to the decay of a conventional $q\bar{q}$ state into two mesons by the quark pair creation (QPC) mechanism. In the flux tube scenario it corresponds to the flux excitation mode along the displacement between the quark and anti-quark, for which the potential is denoted as $\hat{V}_L$.  The transition of Fig.~\ref{fig-1} (b) corresponds to the flux excitation mode transverse to the displacement between the quark and anti-quark. The quark pair created from this mode will recoil the initial color-octet $q\bar{q}$ via the transverse flux motion. For a conventional $q\bar{q}$ decay via the $^3P_0$ QPC mechanism, the kinematic regime as Fig.~\ref{fig-1} (b) will be relatively suppressed with respect to Fig.~\ref{fig-1} (a). Since in such a case, in order to balance the color, an additional relatively-hard gluon will be exchanged between the recoiled $q\bar{q}$ and the created $q\bar{q}$. In contrast, such a transition in the hybrid decay can naturally occur via the transverse mode of the flux tube oscillations~\cite{Kokoski:1985is}. Namely, the created $q\bar{q}$ can easily get the color balanced by soft gluon exchanges which can be absorbed into the effective potential without suppression. Such a transition through the transverse mode of the flux tube motions can be parametrized by the effective potential $\hat{V}_T$.

The transition amplitude for a $1^{-+}$ hybrid of $q\bar{q}\tilde{g}$ decaying into two pseudoscalar mesons can then be expressed as
\begin{eqnarray}
{\cal M}_a &= & \langle (q_1\bar{q}_4)_{M_1}(q_3\bar{q}_2)_{M_2}|\hat{V}_L|q_1\bar{q}_2\tilde{g}\rangle\equiv  g_1 |{\bf k}| \ ,
\end{eqnarray}
and 
\begin{eqnarray}
{\cal M}_b &=& \langle (q_1\bar{q}_2)_{M_1}(q_3\bar{q}_4)_{M_2}|\hat{V}_T|q_1\bar{q}_2\tilde{g}\rangle\equiv  g_2 |{\bf k}| \ ,
\end{eqnarray}
for these two decay modes, respectively. In the above two equations, ${\bf k}$ is the three-vector momentum of the final-state meson in the c.m. frame of the hybrid, and the quarks (anti-quarks) are the non-strange quarks (anti-quarks). Note that the QPC only contributes to a flavor singlet $\tilde{g}\to (u\bar{u}+d\bar{d}+s\bar{s})/\sqrt{3}$. We mention that when the $s\bar{s}$ pair is created, an SU(3) flavor symmetry breaking parameter will be included. Also, in the above two amplitudes the interchanges of the final-state hadron indices are implied. 

This parametrization leads to a connection among the couplings of an initial hybrid state to different SU(3) channels, and they are collected in Table~\ref{tab-1}. Interesting features with the hybrid nonet decays can be learned as follows:

\begin{itemize}

\item It is rather clear that if the final states do not contain isoscalar mesons, the transitions will be via the string breaking potential $\hat{V}_L$ along the displacement between the quark and anti-quark. Namely, the transitions are similar to the conventional $^3P_0$ process. For $K^*$ decays into $K\pi$, it will be difficult to distinguish it from the conventional $q\bar{q}$ vector mesons. 

\item For the $\pi_1$ and $K^*$ decays into $\eta$ or $\eta'$ plus a $I\neq 0$ state, such as $\pi_1^0\to \eta\pi^0$ and $\eta'\pi^0$, the couplings involve interferences between processes of Fig.~\ref{fig-1} (a) and (b). Since the mixing angle between $\eta$ and $\eta'$ is $\alpha_P\simeq 42^\circ$, the couplings for the channels between $\eta$ and $\eta'$ would be very different.  

\item $\eta_{1L}$ and $\eta_{1H}$ decays into $\pi\pi$ and $K\bar{K}$ are forbidden by the Bose symmetry and $G$-parity conservation. They can only access $\eta\eta'$ via the octet and singlet mixing. The coupling strengths have non-trivial dependence of the mixing angle $\alpha$. One can see that the decay pattern for these channels in a combined analysis should be sensitive to the value of $\alpha$. 

\end{itemize}

\begin{table}
\centering
\caption{The coupling constants for the $1^{-(+)}$ hybrid nonet decays into pseudoscalar meson pairs. The couplings for the negative charge states are implied. The SU(3) flavor symmetry breaking parameter $R$ is also included.}
\begin{tabular}{l||c}
\hline 
Processes & Couplings\\
\hline
$\pi_1^0\to \eta\pi^0$ & $\frac{1}{\sqrt{2}}(g_1+g_2)\cos\alpha_P-R g_2\sin\alpha_P$ \\
$\pi_1^0\to \eta'\pi^0$ & $\frac{1}{\sqrt{2}}(g_1+g_2)\sin\alpha_P+R g_2\cos\alpha_P$ \\
$\pi_1^+\to\eta\pi^+$ & $\sqrt{2}(g_1+g_2)\cos\alpha_P-R g_2\sin\alpha_P$ \\
$\pi_1^+\to\eta'\pi^+$ & $\sqrt{2}(g_1+g_2)\sin\alpha_P+R g_2\cos\alpha_P$ \\
$\eta_{1L}\to \eta\eta'$ & $\frac{1}{2}(g_1+g_2)\sin 2\alpha_P(\cos\alpha+R\sin\alpha)+g_2\cos 2\alpha_P(R\cos\alpha-\sin\alpha)$ \\
$\eta_{1H}\to \eta\eta'$ & $\frac{1}{2}(g_1+g_2)\sin 2\alpha_P(\sin\alpha-R\cos\alpha)+g_2\cos 2\alpha_P(R\sin\alpha+\cos\alpha)$ \\
$K^{*+}\to K^+\pi^0$ & $\frac{1}{\sqrt{2}} g_1$ \\
$K^{*+}\to K^0\pi^+$ & $g_1$ \\
$K^{*+}\to K^+\eta$ & $g_1(\frac{1}{\sqrt{2}}\cos\alpha_P- R\sin\alpha_P) + g_2(\sqrt{2}\cos\alpha_P- R\sin\alpha_P)$ \\
$K^{*+}\to K^+\eta'$ & $g_1(\frac{1}{\sqrt{2}}\sin\alpha_P+ R\cos\alpha_P) + g_2(\sqrt{2}\sin\alpha_P+ R\cos\alpha_P)$ \\
\hline 
\end{tabular}
\label{tab-1}
\end{table}

\subsection{$J/\psi\to\gamma \eta_1\to\gamma \eta\eta'$ }

\begin{figure}
\includegraphics[width=2.5in]{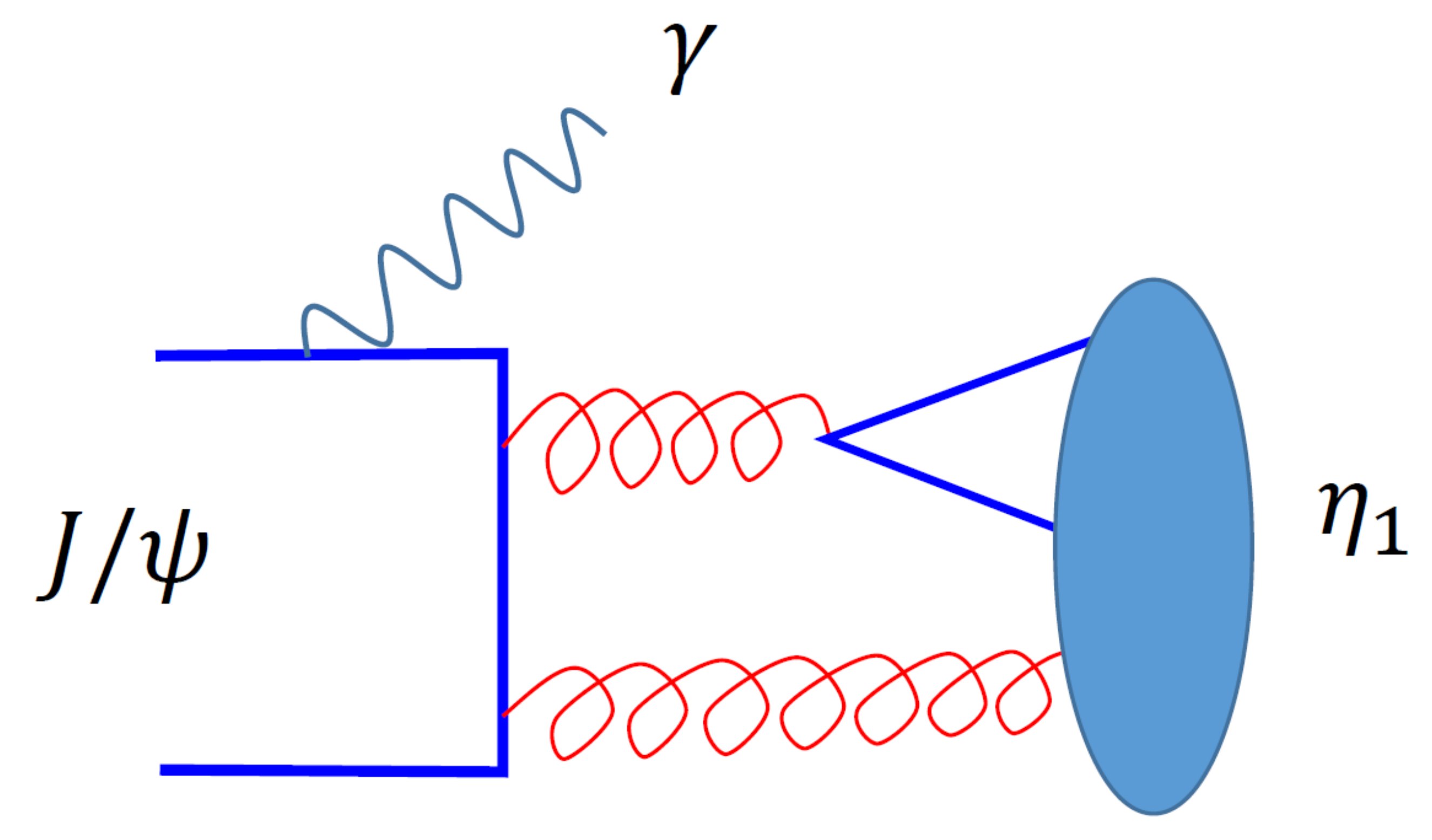}
\caption{Illustration of the $1^{-+}$ isoscalar hybrid production in $J/\psi\to \gamma \eta_1$.}
\label{fig-2}
\end{figure}

A typical process for the production of a $J^{PC}=1^{-+}$ hybrid in the $J/\psi$ radiative decays is illustrated by Fig.~\ref{fig-2}. It shows that the annihilations of the charm and anti-charm quark can create a pair of light $S$-wave $q\bar{q}$  associated by a constituent gluon in a relative $P$-wave to the $q\bar{q}$.  
At the hadronic level, the Lagrangian for a general vector-vector-vector field interaction at the leading-order can be described by
\begin{equation}
  \label{eq:vvv}
  \mathcal{L}_{VVV}=ig_{VVV}(V_{1,\nu}\overleftrightarrow{\partial^{\mu}}V_2^{\nu}V_{3,\mu}+V_{1,\mu}V_2^{\nu}\overleftrightarrow{\partial^{\mu}}V_{3,\nu}+V_{2,\mu}V_3^{\nu}\overleftrightarrow{\partial^{\mu}}V_{1,\nu})\ ,
\end{equation}
where $V_1$, $V_2$, $V_3$ denotes the vector fields. For the radiative decay of $J/\psi\to\gamma \eta_1$, since the photon is transversely polarized, the above Lagrangian will reduce to the following form:
\begin{equation}
{\cal L}_{J/\psi\to \gamma \eta_1} = ig_{J/\psi \eta_1\gamma}F_{\mu\nu}V_{J/\psi}^\mu V_{\eta_1}^\nu \ ,
\end{equation}
where $F_{\mu\nu}\equiv \partial_\mu A_\nu-\partial_\nu A_\mu$, and the vector fields $V_{J/\psi}$, $A$, and $V_{\eta_1}$ stand for the initial $J/\psi$, final-state photon, and hybrid $\eta_1$ fields, respectively; $g_{J/\psi \eta_1\gamma}$ is the coupling constant. Note that the leading transition of $J/\psi\to\gamma \eta_1$ is via a $P$ wave. In the center of mass (c.m.) frame of $J/\psi$ the squared transition amplitudes for the two $I=0$ states can be expressed like below:
\begin{eqnarray}
|i{\cal M}(J/\psi\to\gamma \eta_{1L})|^2&\propto&  g_{J/\psi \eta_{1L}\gamma}^2|{\bf q}_L|^2(1+m_{J/\psi}^2/m_{\eta_{1L}}^2) \ , \\
|i{\cal M}(J/\psi\to\gamma \eta_{1H})|^2&\propto&  g_{J/\psi \eta_{1H}\gamma}^2|{\bf q}_H|^2(1+m_{J/\psi}^2/m_{\eta_{1H}}^2)\ , \\
\end{eqnarray}
where ${\bf q}_L$ and ${\bf q}_H$ are the three-vector momenta of $\eta_{1L}$ and $\eta_{1H}$ in the $J/\psi$ rest frame, respectively. The subscripts, ``$L$" and ``$H$", stand for the low and high mass states, respectively. The two coupling constants, $g_{J/\psi \eta_{1L}\gamma}$ and $g_{J/\psi \eta_{1H}\gamma}$, which account for the production mechanism for these two isoscalars, can be parametrized out:
\begin{eqnarray}\label{prod-coupling}
g_{J/\psi \eta_{1L}\gamma} &=& g_0(\sqrt{2}\cos\alpha-R\sin\alpha) \ ,\\
g_{J/\psi \eta_{1H}\gamma} &=& g_0(\sqrt{2}\sin\alpha+R\cos\alpha) \ ,
\end{eqnarray}
where $R\simeq f_\pi/f_K\simeq 0.93$ indicates the SU(3) flavor symmetry breaking effects in the production of the $s\bar{s}$ pair in comparison with the non-strange $q\bar{q}$ pairs, and $g_0$ describes the coupling strength for the production of a light hybrid configuration $q\bar{q}\tilde{g}$ of $J^{PC}=1^{-+}$ in the $J/\psi$ radiative decays. It can be expressed as
\begin{equation}\label{def-coup}
g_0\equiv \langle (q\bar{q}\tilde{g})_{1^{-+}} |\hat{H}_{em}|J/\psi\rangle \ ,
\end{equation}
where $\hat{H}_{em}$ contains the dynamics for the transition of Fig.~\ref{fig-2}. 

The coupling relation in Eq.~(\ref{prod-coupling}) leads to the relative production rate  for $\eta_{1L}$ and $\eta_{1H}$ as follows:
\begin{eqnarray}\label{prod-rate-LH}
r_{L/H}&\equiv &\frac{BR(J/\psi\to\gamma \eta_{1L})}{BR(J/\psi\to\gamma \eta_{1H})}= \left(\frac{|{\bf q}_L|}{|{\bf q}_H|}\right)^3\frac{(\sqrt{2}\cos\alpha-R\sin\alpha)^2}{(\sqrt{2}\sin\alpha+R\cos\alpha)^2}\frac{m_{\eta_{1H}}^2(m_{J/\psi}^2+m_{\eta_{1L}}^2)}{m_{\eta_{1L}}^2(m_{J/\psi}^2+m_{\eta_{1H}}^2)} \ ,
\end{eqnarray}
which seems to be sensitive to the mixing angle $\alpha$. 
Note that, in Ref.~\cite{besiii-hybrid,besiii-hybrid-pwa} the $\eta_1(1855)$ signal is actually observed in its decays into $\eta\eta'$. Moreover, the PWA results suggest that only one $I=0$ hybrid state has been clearly seen in the $1^{-+}$ partial wave amplitude. As shown in Subsection~\ref{subsect-nonet-decay}, the decays of $\eta_{1L}$ and $\eta_{1H}$ into $\eta\eta'$ are strongly correlated with the mixing angle $\alpha$ and mechanisms for the flux tube breaking. It means that the following branching ratio fractions can serve as constraints on the mixing angle:
\begin{eqnarray}\label{scheme-1}
\textbf{Scheme-I} & : & R_{\eta_{1L}/\eta_1(1855)}\equiv \frac{BR(J/\psi\to\gamma \eta_{1L}\to\gamma\eta\eta')}{BR(J/\psi\to\gamma \eta_1(1855)\to\gamma\eta\eta')} < 10 \% \ ,
\end{eqnarray}
and 
\begin{eqnarray}\label{scheme-2}
\textbf{Scheme-II} & : & R_{\eta_{1H}/\eta_1(1855)}\equiv \frac{BR(J/\psi\to\gamma \eta_{1H}\to\gamma\eta\eta')}{BR(J/\psi\to\gamma \eta_1(1855)\to\gamma\eta\eta')} < 10 \% \ ,
\end{eqnarray}
where we have assigned $\eta_1(1855)$ as either the higher mass state (Scheme-I) or the lower mass state (Scheme-II). The relative rate $10\%$ is the production upper limit for the partner of $\eta_1(1855)$ from the experimental measurement~\cite{besiii-hybrid,{besiii-hybrid-pwa}}.

With the production and decay couplings extracted earlier, the general form for the joint branching ratio fraction can be  expressed as
\begin{eqnarray}\label{joint-prod-decay}
R_{\eta_{1L}/\eta_{1H}}&=&\left(\frac{|{\bf q}_L|}{|{\bf q}_H|}\right)^3\frac{(\sqrt{2}\cos\alpha-R\sin\alpha)^2}{(\sqrt{2}\sin\alpha+R\cos\alpha)^2}\frac{m_{\eta_{1H}}^2(m_{J/\psi}^2+m_{\eta_{1L}}^2)}{m_{\eta_{1L}}^2(m_{J/\psi}^2+m_{\eta_{1H}}^2)}  \left(\frac{|{\bf k}_L|}{|{\bf k}_H|}\right)^3 \left(\frac{\Gamma_Hm_{\eta_{1H}}}{\Gamma_Lm_{\eta_{1L}}}\right)^2 \nonumber\\
&\times & \frac{[(1+\delta)\tan 2\alpha_P(\cos\alpha+R\sin\alpha)+2 \delta(R\cos\alpha-\sin\alpha)]^2}{[(1+\delta)\tan 2\alpha_P(\sin\alpha-R\cos\alpha)+ 2\delta(R\sin\alpha+\cos\alpha)]^2} \ ,
\end{eqnarray}
where $\Gamma_L$ and $\Gamma_H$ are the total widths of the lower and higher mass states, respectively; ${\bf q}_{L,H}$ and ${\bf k}_{L,H}$ are the three-vector momenta of the photon and pseudoscalar meson in the rest frames of $J/\psi$ and $\eta_{L,H}$, respectively; $\delta\equiv g_2/g_1$ indicates the relative strength between the two decay mechanisms for the flux tube breaking. As discussed earlier, $|\delta|\simeq 1$ is for the hybrid decays, while $|\delta|<< 1$ for conventional $q\bar{q}$ decays. In Eq.~(\ref{joint-prod-decay}) if we approximate ${\Gamma_H}/{\Gamma_L}\simeq 1$, ratio $R_{\eta_{1L}/\eta_{1H}}$ will strongly depend on $\alpha$ and $\delta$.

\subsection{Results and analyses}

Before we go to the detailed studies of the two schemes, we briefly summarize the present experimental information on the  strange vector mesons. As listed by the Particle Data Group (PDG)~\cite{ParticleDataGroup:2020ssz}, two excited $K^*$ states are observed in experiment, i.e. $K^*(1410)$ and $K^*(1680)$. While $K^*(1410)$ can be well accommodated by the  first radial excitations of the vector meson nonet, the property of $K^*(1680)$ is far from well explored. Note that the second radial excitations of the isoscalar pseudoscalar mesons can be occupied by $\eta(1760)$ and $\eta(1860)$ in the Regge trajectory~\cite{Yu:2011ta}, the mass of $K^*(1680)$ as the second radial excitation in the conventional $q\bar{q}$ vector nonet seems to be too small. We also note that the strange pseudoscalar partner in the second radial excitation nonet has not yet been established in experiment though $K(1630)$ could be a candidate~\cite{ParticleDataGroup:2020ssz}. In the following analysis we first treat $K^*(1680)$ as the strange partner of the $1^{-(+)}$ nonet and examine whether it fits the constraint. If not, we then investigate the mass correlation of $K^*$ with the mixing angle and other multiplets as required by the Gell-Mann-Okubo relation.

\subsubsection{Scheme-I}

\begin{figure}
\includegraphics[width=2.5in]{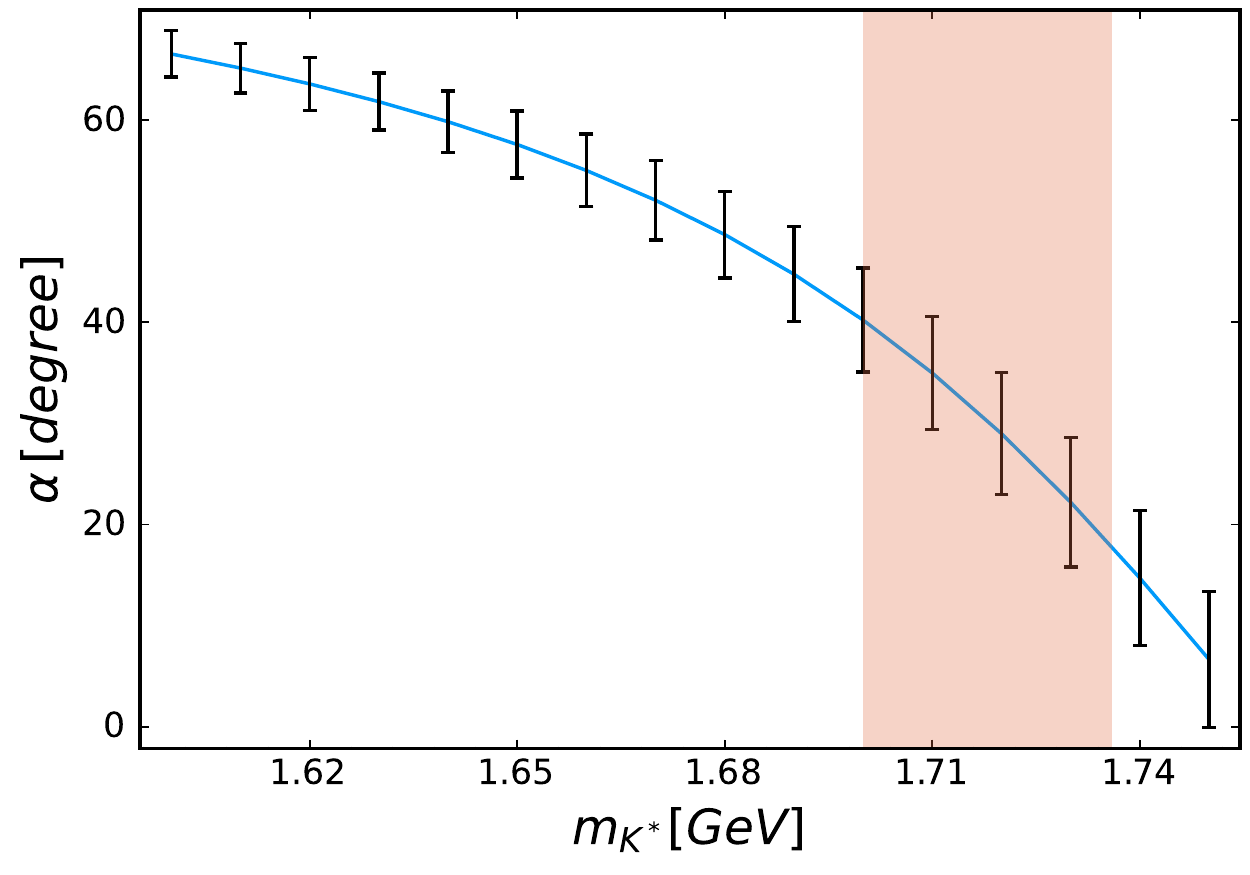}
\caption{The correlation of the mixing angle $\alpha$ with the $K^*$ mass. The uncertainties are due to the mass uncertainties for those input states. The shadowed area indicates the mass range of $K^*$ from PDG~\cite{ParticleDataGroup:2020ssz}.}
\label{fig-alpha-kstar}
\end{figure}

With the $\eta_1(1855)$ assigned as the higher mass state, and $\pi_1(1600)$ and $K^*(1680)$ assigned as the $I=1$ and strange partner, respectively, we extract $m_{\eta_{1L}}=1712.5\pm 8.7$ MeV from Eq.~(\ref{GMO-mass}), and the mixing angle $\alpha=30^\circ\pm 13^\circ$. The uncertainties are given by the mass uncertainties from $\pi_1(1600)$ and $K^*(1680)$ via the Gell-Mann-Okubo relation. We note that the PDG values~\cite{ParticleDataGroup:2020ssz} are adopted for the masses of these two states, i.e. $m_{\pi_1}=1661\pm 13$ MeV and $m_{K^*}=1718\pm 18$ MeV. From Eq.~(\ref{gell-mann-okubo-rela}), we can extract the correlation between the mixing angle and the $K^*$ mass, and the results are presented in Fig.~\ref{fig-alpha-kstar}.

Although the uncertainties of the mixing angle $\alpha$ seem to be rather large, it indicates significant mixings between the flavor octet and singlet, and apparently deviates from the ideal mixing. This seems to be a necessary consequence if there is only one $I=0$ hybrid state observed in the $\eta\eta'$ channel in $J/\psi\to\gamma\eta\eta'$. Nevertheless, it favors the hybrid scenario to have important contributions from the transverse mode of the flux tube motions. To illustrate this, we first look at Eq.~(\ref{prod-rate-LH}) where by taking the limit of ideal mixing, i.e. $\alpha=0^\circ$, the production ratio $r_{L/H}\simeq 2$ can be obtained. Note that the ratio $r_{L/H}$ is insensitive to the phase space factor and the SU(3) flavor symmetry breaking parameter $R$. 

In the case that $\eta_1(1855)$ is the higher mass state, it is the ratio $R_{\eta_{1L}/\eta_{1H}}$ defined in Eq.~(\ref{joint-prod-decay}) can be compared with the experimental observables with $R_{\eta_{1L}/\eta_1(1855)}<10\%$. In the ideal mixing limit, one has 
\begin{eqnarray}
R_{\eta_{1L}/\eta_1(1855)}&\simeq &\left(\frac{|{\bf q}_L|}{|{\bf q}_H|}\right)^3\left(\frac{|{\bf k}_L|}{|{\bf k}_H|}\right)^3\frac{m_{\eta_{1H}}^2(m_{J/\psi}^2+m_{\eta_{1L}}^2)}{m_{\eta_{1L}}^2(m_{J/\psi}^2+m_{\eta_{1H}}^2)}  \left(\frac{\Gamma_Hm_H}{\Gamma_Lm_L}\right)^2  \frac{2}{R^2}\left[\frac{(1+\delta)\tan 2\alpha_P+2R\delta}{R(1+\delta)\tan 2\alpha_P-2\delta}\right]^2
 \ .
\end{eqnarray}
Note that the product $\left(\frac{|{\bf q}_L|}{|{\bf q}_H|}\right)^3\left(\frac{|{\bf k}_L|}{|{\bf k}_H|}\right)^3\frac{m_{\eta_{1H}}^2(m_{J/\psi}^2+m_{\eta_{1L}}^2)}{m_{\eta_{1L}}^2(m_{J/\psi}^2+m_{\eta_{1H}}^2)}  \left(\frac{\Gamma_Hm_H}{\Gamma_Lm_L}\right)^2$ actually enhances the ratio, and $\tan 2\alpha_P\simeq 10$ will further push up the ratio. It thus relies on the value of $\delta$ to decide the value of $R_{\eta_{1L}/\eta_{1H}}$. As discussed earlier, for conventional $q\bar{q}$ meson decays, one would expect $\delta\to 0$. It actually leads to $R_{\eta_{1L}/\eta_{1H}}>1$ which is in contradiction with the experimental observation. For the hybrid decays, the transverse mode of the flux tube motions plays an important role in the decays. It means $|\delta|\simeq 1$, or even to be the dominant transition mechanism with $|\delta|>1$. Eventually,  to have $R_{\eta_{1L}/\eta_1(1855)}<10\%$ as suggested by the experimental data, one finds $\delta$ should take a negative value and the absolute value is at ${\cal O}(1)$. 

\begin{figure}
\includegraphics[width=4.5in]{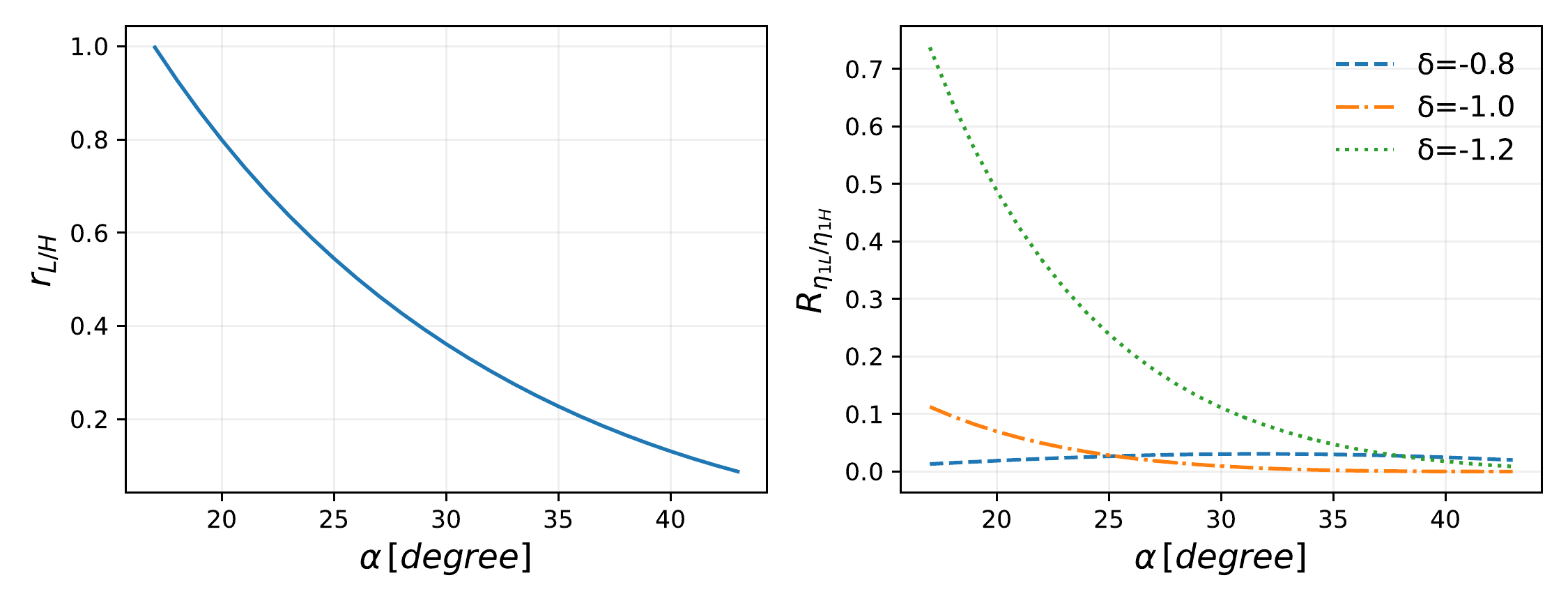}
\caption{The dependence of the ratios  $r_{L/H}$ and $R_{\eta_{1L}/\eta_1(1855)}$ on the mixing angle $\alpha$ within the preferred range of $\alpha\in (17^\circ, \ 43^\circ)$. In Scheme-I $\eta(1855)$ is the high-mass state. On the left panel the solid line is for $r_{L/H}$, while on the right panel the dashed, dot-dashed, and dotted lines correspond to the ratio $R_{\eta_{1L}/\eta_1(1855)}$ with $\delta=-0.8, \ -1.0, \ -1.2$, respectively. }
\label{fig-ratio-in-alpha}
\end{figure}

In Fig.~\ref{fig-ratio-in-alpha}, with the mixing angle $\alpha$ within its uncertainty range, i.e. $\alpha=30^\circ\pm 13^\circ$, we plot the ratios  $r_{L/H}$ and $R_{\eta_{1L}/\eta_1(1855)}$. For demonstration, we adopt $\delta=-0.8, \ -1.0, \ -1.2$ to calculate $R_{\eta_{1L}/\eta_1(1855)}$. It shows that $r_{L/H}$ is not sufficiently suppressed while the decays via the transverse mode play a dominant role to suppress the low-mass state. Although we cannot give a precise value for $\delta$ based on the present experimental results, we find that the relative sign between $g_1$ and $g_2$ and their relative strength can consistently reflect the hybrid features.

\begin{figure}[htbp]
\includegraphics[width=3.5in]{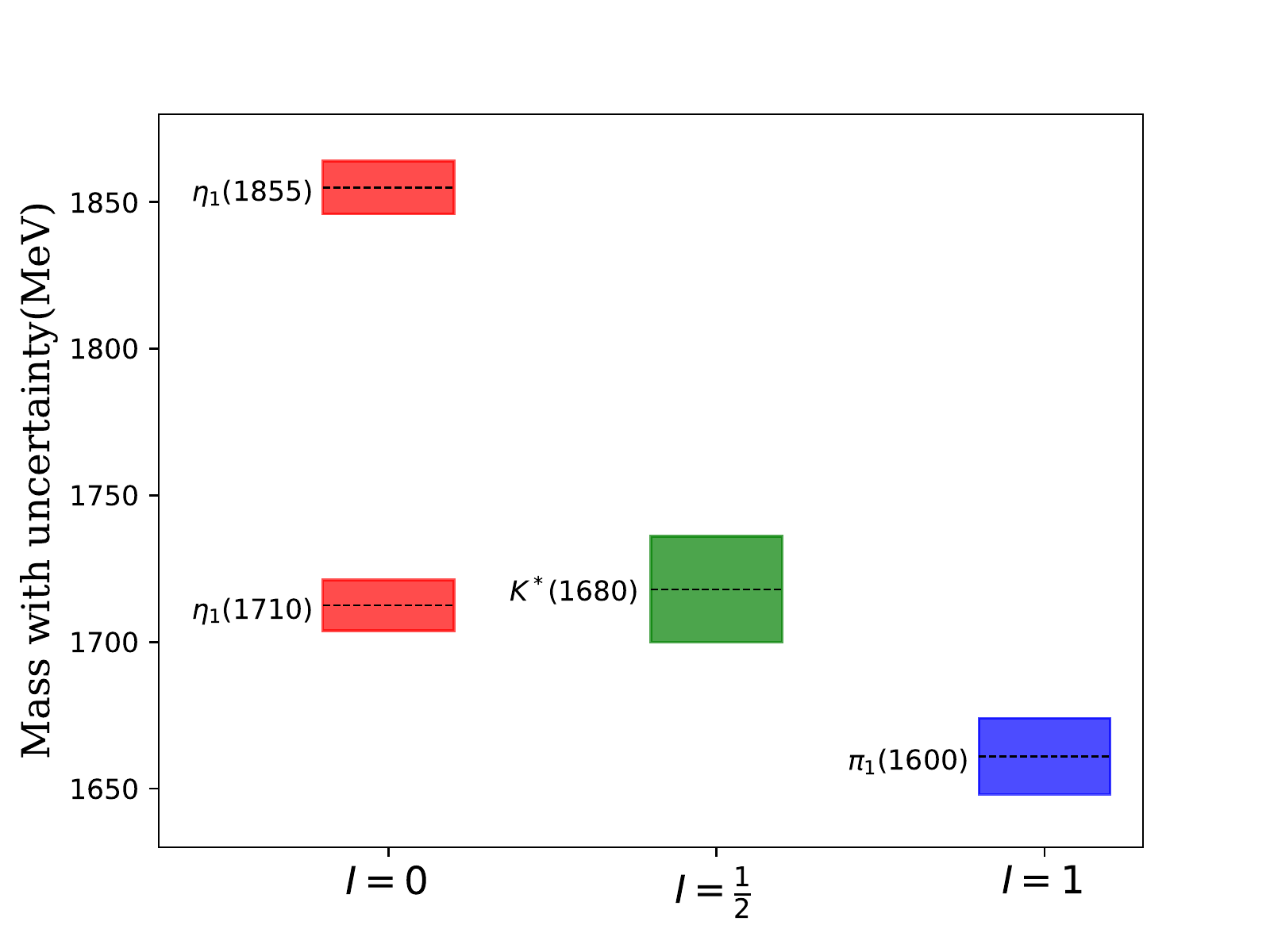}
\caption{The $1^{-(+)}$ hybrid nonet with mass uncertainties determined in Scheme-I. Namely, $\eta_1(1855)$ is assigned as the higher mass state with $I=0$. }
\label{fig-spectrum-1}
\end{figure}

In Fig.~\ref{fig-spectrum-1}, we illustrate the $1^{-(+)}$ nonet in Scheme-I. The shadowed ranges are the mass uncertainties and the central dashed lines denote the preferred mass.

\subsubsection{Scheme-II}

With the $\eta_1(1855)$ assigned as the lower mass state, and keep $\pi_1(1600)$ as the $I=1$ partner, the determination of the lower $I=0$ state will be correlated with the mass of the strange partner differently. It means that $K^*(1680)$ is no longer suitable for being the strange partner of the nonet. This can be seen easily via Eq.~(\ref{GMO-mass}) which is symmetric to $\eta_{1L}$ and $\eta_{1H}$. If the same $K^*$ mass is taken, the solution for the other $\eta_1$ mass will be a lower one as in Scheme-I, and $\eta_1(1855)$ will keep to be the higher mass state. 

Searching for the higher mass partner of $\eta_1(1855)$ thus needs a higher $K^*$ mass as input. As discussed earlier, so far we still lack experimental information about the vector strange spectrum. Fortunately, if we impose again the BESIII observation as a constraint, we should have the inverse form of Eq.~(\ref{joint-prod-decay}) to satisfy Eq.~(\ref{scheme-2}). In such a case, we find that the mixing angle is still located around $\alpha\in (25^\circ, 45^\circ)$ corresponding to $m_{K^*}\simeq 1.83\sim 1.90$ GeV. Meanwhile, it shows that $\delta$ is still at ${\cal O}(1)$, but favors a positive sign. In fact, the sign and magnitude of $\delta$ turn out to be very sensitive to the experimental constraint which can be seen analytically via Eq.~(\ref{joint-prod-decay}).

\begin{figure}
\includegraphics[width=2.5in]{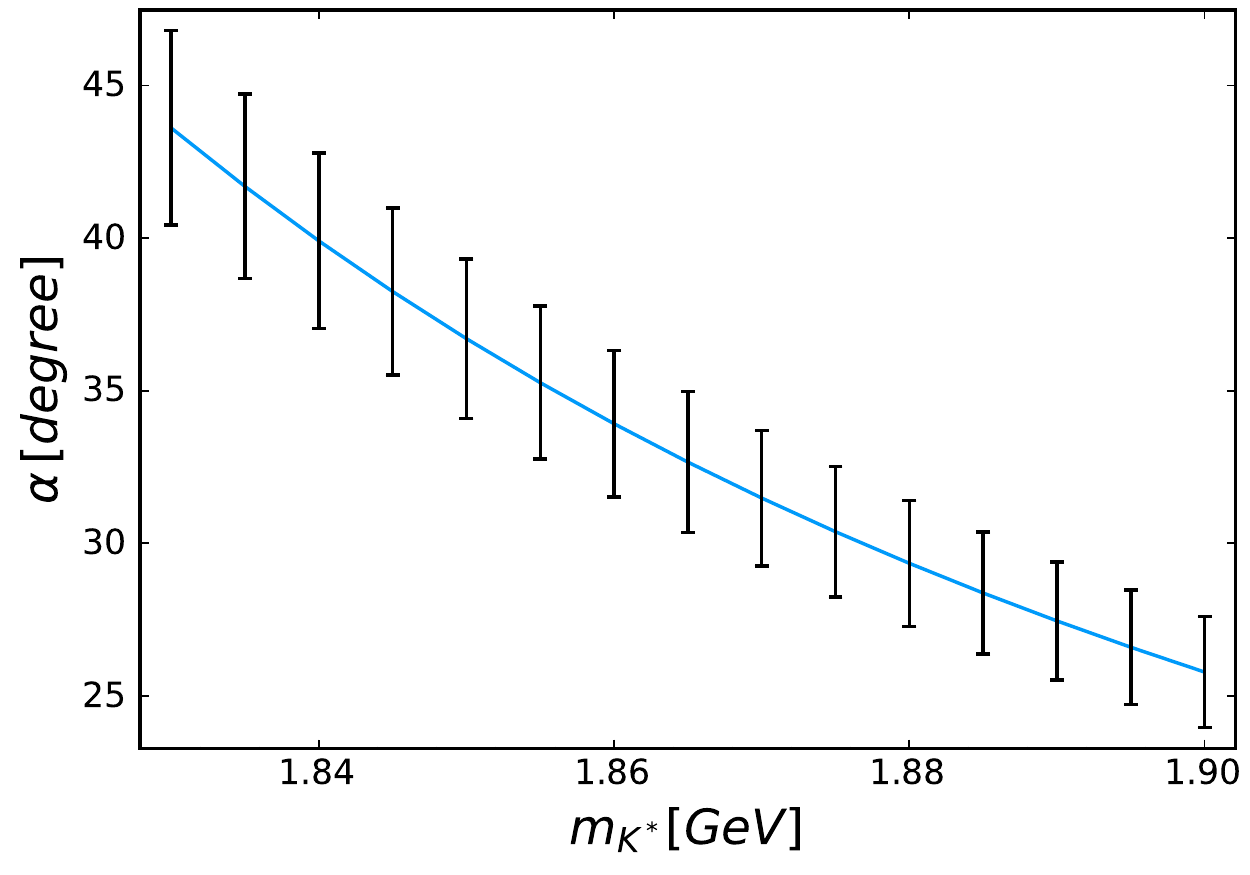}
\caption{The correlation of the mixing angle $\alpha$ with the $K^*$ mass in Scheme-II. The uncertainties are due to the mass uncertainties for those input states.}
\label{fig-alpha-kstar-high}
\end{figure}

Similar to Fig.~\ref{fig-alpha-kstar}, we plot in Fig.~\ref{fig-alpha-kstar-high} the correlation between the  mixing angle $\alpha$ and the $K^*$ mass. The preferred $K^*$ mass is $m_{K^*}\simeq 1.83\sim 1.90$ GeV corresponding to the range of $\alpha=25^\circ\sim 45^\circ$.

\begin{figure}
\includegraphics[width=4.5in]{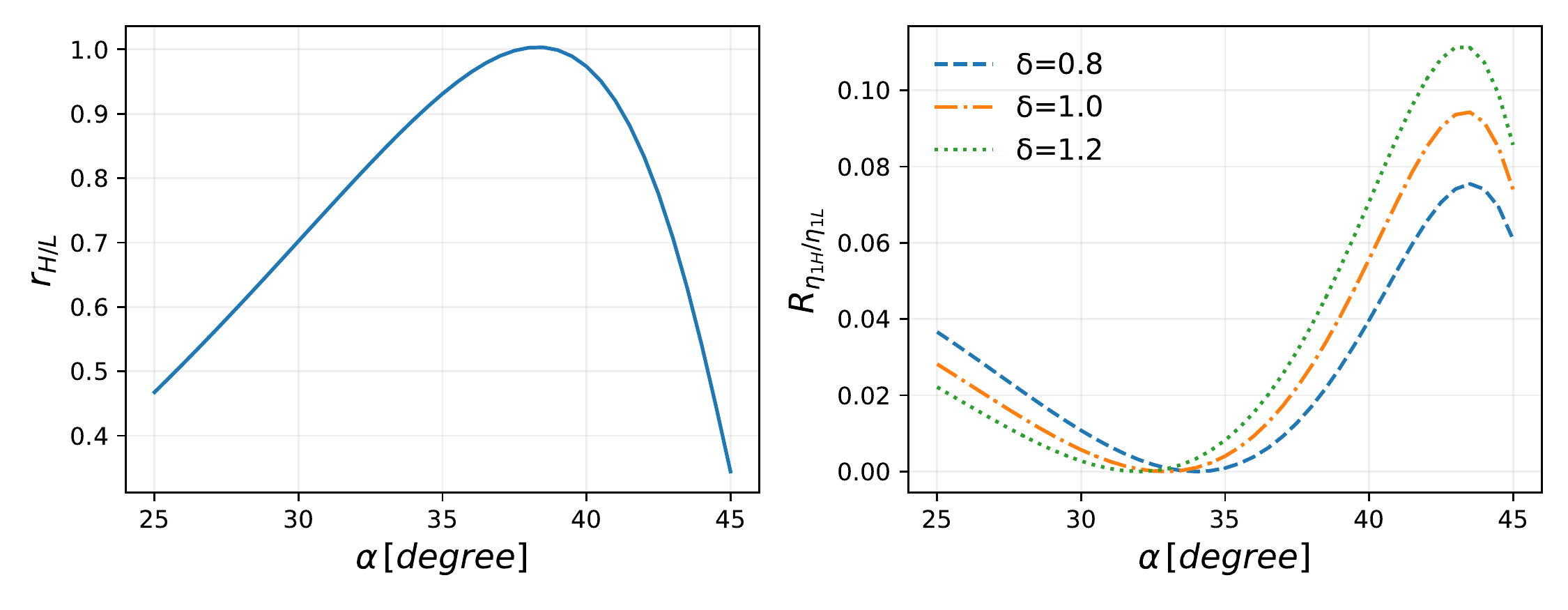}
\caption{The dependence of the ratios  $r_{H/L}$ and $R_{\eta_{1H}/\eta_1(1855)}$ on the mixing angle $\alpha$ within the preferred range of $\alpha\in (25^\circ, \ 45^\circ)$. In Scheme-II, $\eta_1(1855)$ is the low-mass state. On the left panel the solid line is for $r_{H/L}$, while on the right panel the dashed, dot-dashed, and dotted lines correspond to the ratio $R_{\eta_{1H}/\eta_1(1855)}$ with $\delta=0.8, \ 1.0, \ 1.2$, respectively.  }
\label{fig-ratio-in-alpha-2}
\end{figure}

In Fig.~\ref{fig-ratio-in-alpha-2}, we present the results for $r_{H/L}$ and $R_{\eta_{1H}/\eta_1(1855)}$ in terms of  the mixing angle $\alpha$ within its uncertainty range, i.e. $\alpha= 25^\circ\sim 45^\circ$. It is interesting to see that in Scheme-II the production of $\eta(1855)$ in $J/\psi\to \gamma \eta_{1L, 1H}$ as the low-mass state is actually comparable with the higher one. It is the decay transition of $\eta_{1L}\to\eta\eta'$ that strongly enhances the signal of $\eta_1(1855)$ in the $\eta\eta'$ channel while the higher mass state is suppressed due to its weak coupling to the $\eta\eta'$ channel. Again, we see the dominance of the transverse mode in the hybrid decays. For demonstration, we adopt $\delta=0.8, \ 1.0, \ 1.2$ to calculate $R_{\eta_{1H}/\eta_1(1855)}$ which are presented on the right panel of Fig.~\ref{fig-ratio-in-alpha-2}.

\begin{figure}[htbp]
\includegraphics[width=3.5in]{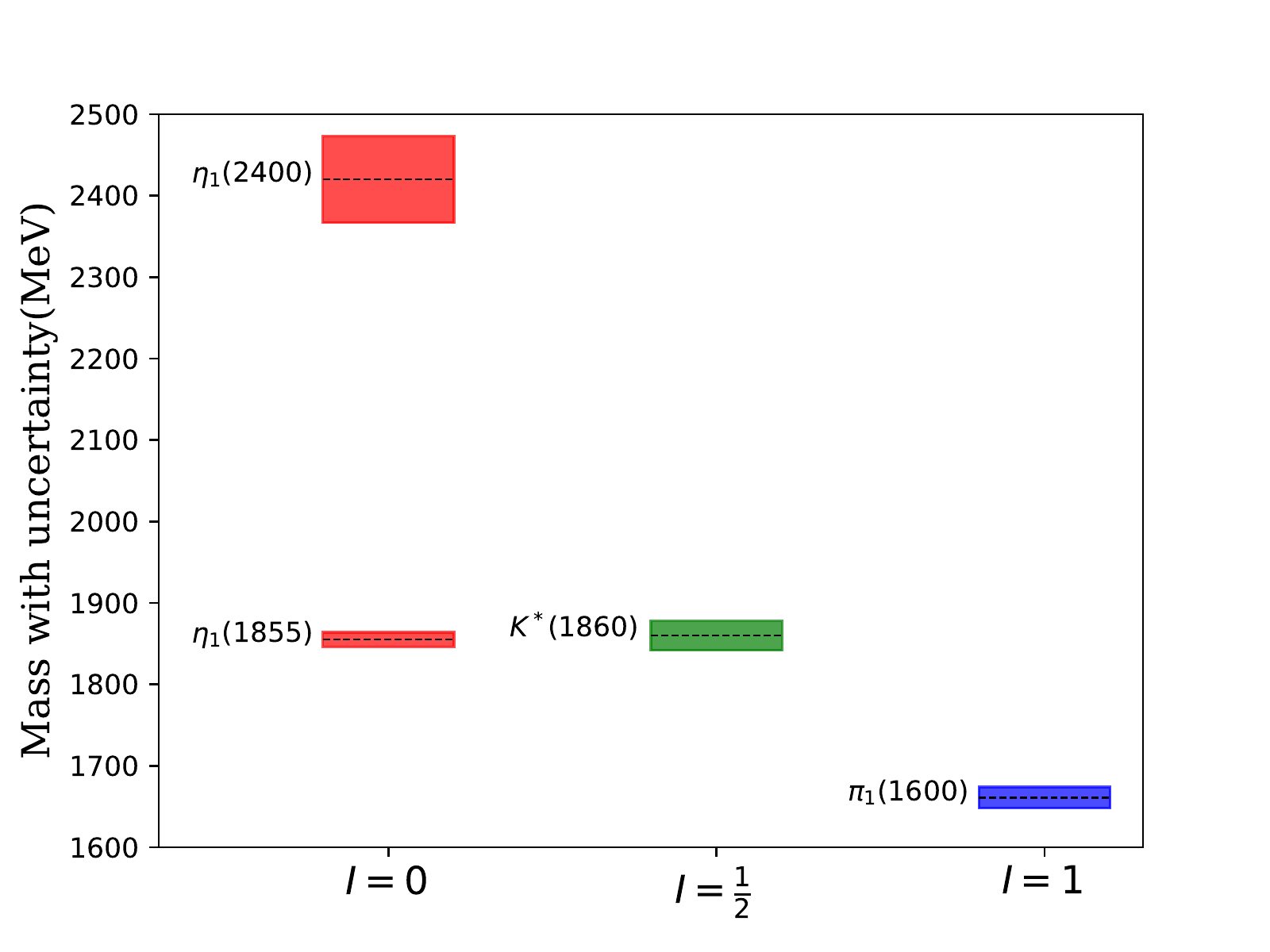}
\caption{The $1^{-(+)}$ hybrid nonet with mass uncertainties determined in Scheme-II. Namely, $\eta_1(1855)$ is assigned as the lower mass state with $I=0$. }
\label{fig-spectrum-2}
\end{figure}

In Fig.~\ref{fig-spectrum-2}, we illustrate the $1^{-(+)}$ nonet in Scheme-II. The shadowed ranges are the mass uncertainties and the central dashed lines denote the preferred mass. 

Comparing these two nonet schemes, it shows that the transverse mode plays an important role for understanding the decay pattern observed in experiment. The relative sign between the transverse mode and the longitudinal mode should decide which scheme is the physical one. Based on the present experimental information, it is however impossible to conclude. We would look forward to further observables to provide a constraint on the sign from experiment. Meanwhile, we note that the LQCD calculations of these two decay modes may also be useful for determining their relative sign.

\subsection{Predictions for $J/\psi\to V H$}

\begin{figure}[htbp]
\includegraphics[width=2.5in]{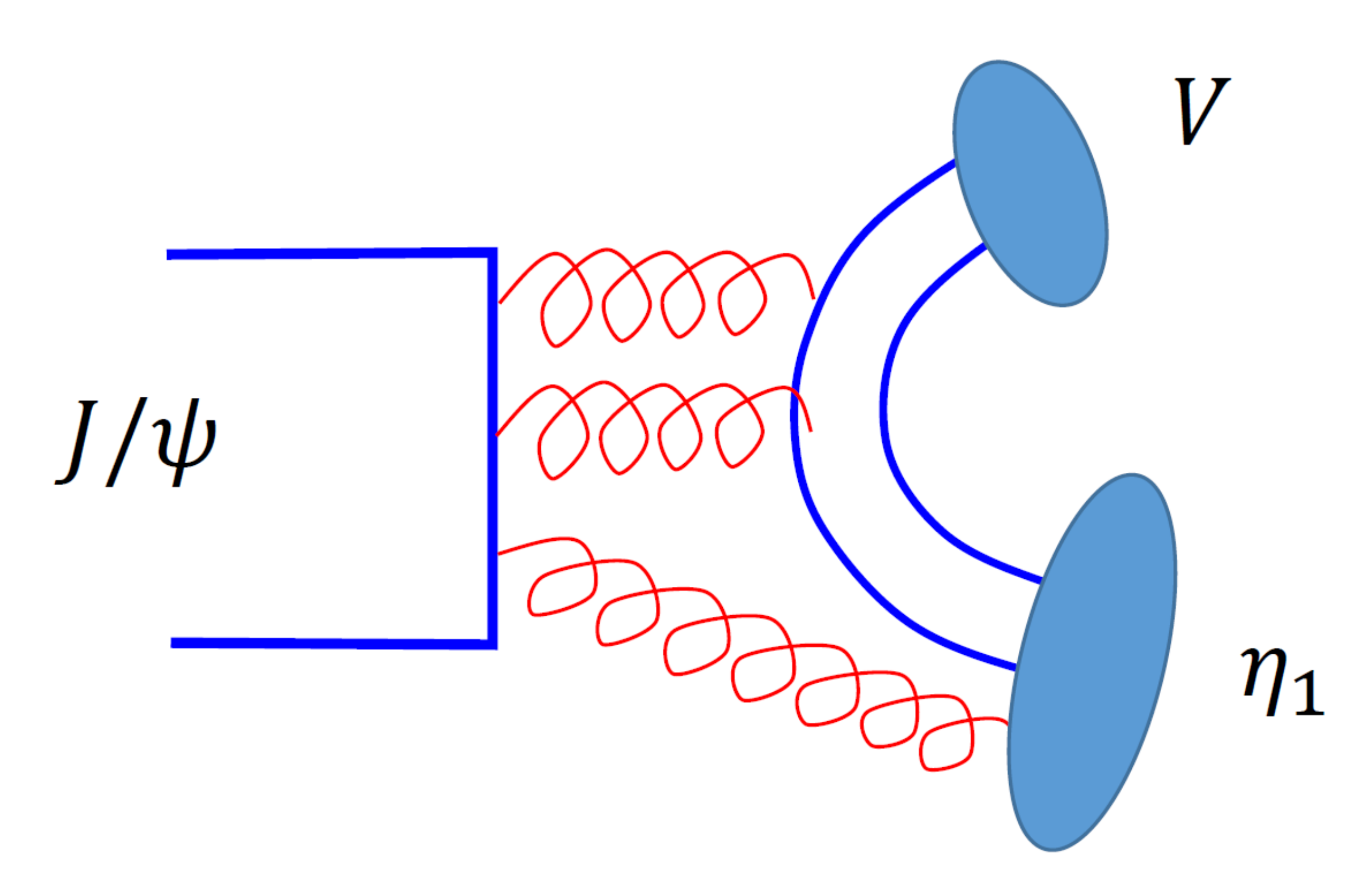}
\caption{Illustration of the production process for the $1^{-(+)}$ hybrid states in $J/\psi\to VH$ where $V$ stands for the light vector mesons $\rho$, $\omega$, $\phi$ and $K^*(892)$. }
\label{fig-prod-VH}
\end{figure}

In order to further investigate the characters arising from the nonet structure of the $1^{-(+)}$ hybrid states, we analyse the hadronic decays of $J/\psi\to V H$ and look for signals for the $I=0$ partner of $\eta_1(1855)$. Here, $V$ stands for the vector mesons $\rho$, $\omega$, and $\phi$, while $H$ stands for the light hybrid multiplets. This process is illustrated by Fig.~\ref{fig-prod-VH} and the leading-order Lagrangian has been given in Eq.~(\ref{eq:vvv}). The coupling for $J/\psi\rightarrow [q\bar{q}]_{1^{--}}[q\bar{q}\tilde{g}]_{1^{-+}}$  can be parametrized as,
\begin{eqnarray}
g_P\equiv \langle [q\bar{q}]_{1^{--}}[q\bar{q}\tilde{g}]_{1^{-+}} |\hat{V}_P|J/\psi\rangle \ ,
\end{eqnarray}
where $\hat{V}_P$ represents the potential for the hadronic decays of $J/\psi\to VH$. And the detailed coupling constants for different decay channels list as following:
\begin{eqnarray}
g_{J/\psi\rho^+\pi_1^-} &=&g_P  ,\nonumber\\
g_{J/\psi\omega\eta_{1L}} &=& g_P\cos\alpha  \ ,\nonumber\\
g_{J/\psi\omega\eta_{1H}}  &=& g_P\sin\alpha   \ ,\nonumber\\
g_{J/\psi\phi\eta_{1L}}  &=& -g_P R^2 \sin\alpha   \ ,\nonumber\\
g_{J/\psi\phi\eta_{1H}} &=& g_P R^2 \cos\alpha   \ ,\nonumber\\
g_{J/\psi K^{*+} K_H^{*-}} &=& g_P R  \ ,
\end{eqnarray}
where $R$ is the SU(3) flavor symmetry breaking factor defined earlier. In this Section to distinguish the hybrid $K^*$ from $K^*(892)$, we denote it as $K^*_H$. 
Apart from the partial wave factor ($\propto |{\bf q}|^3$) which should be included for each channel and a mass function which has the same form for each channel, the branching ratio fractions among all the $VH$ decay channels will be driven by  the following relative strengths:
\begin{eqnarray}
&&\rho^+\pi_1^- : \omega\eta_{1L} : \omega\eta_{1H} : \phi\eta_{1L} : \phi\eta_{1H} : K^{*+} K_H^{*-} \nonumber\\
&=&1 : \cos^2\alpha : \sin^2\alpha : R^4\sin^2\alpha : R^4\cos^2\alpha : R^2 \ .
\end{eqnarray}
Note that for the total of $\rho\pi$, a factor of 3 should be multiplied to the $\rho^+\pi_1^-$ channel, while the total of $K^*\bar{K}_H+c.c.$, a factor of 4 should be multiplies to the $K^{*+} K_H^{*-}$ channel.

If we take into account the effects from the partial wave factor and the SU(3) flavor symmetry breaking factor $R$, we find that the $\rho\pi_1$ channel has the largest branching ratio, while the production strengths for most of the other channels are actually comparable with $\alpha\sim 30^\circ$. In particular, it suggests that the production of $\eta_{1L}$ and $\eta_{1H}$ are accessible in the same channel such as $J/\psi\to \omega\eta\eta'$. This is different from the case of $J/\psi\to\gamma\eta\eta'$.

It should also be interesting to study the $J/\psi\to K^* \bar{K}_H^*+c.c.$ channel. Since $K_H^*$ has $J^P=1^-$ which are the same as the radial excitation states of the vector $K^*(892)$, it is difficult to identify the hybrid-like state. But as discussed earlier, the isolated $K^*(1680)$, either as a radial excitation state of $K^*(892)$ or a hybrid state, will bring crucial understandings of the $K^*$ spectrum. In the hybrid scenario $K_H^*$ will favor to decay into $K_1\pi\to K^*\pi\pi$. It means that $J/\psi\to K\bar{K}\pi\pi\pi$ will be ideal for the search of $K_H^*$.

\begin{figure}[htbp]
\includegraphics[width=3.5in]{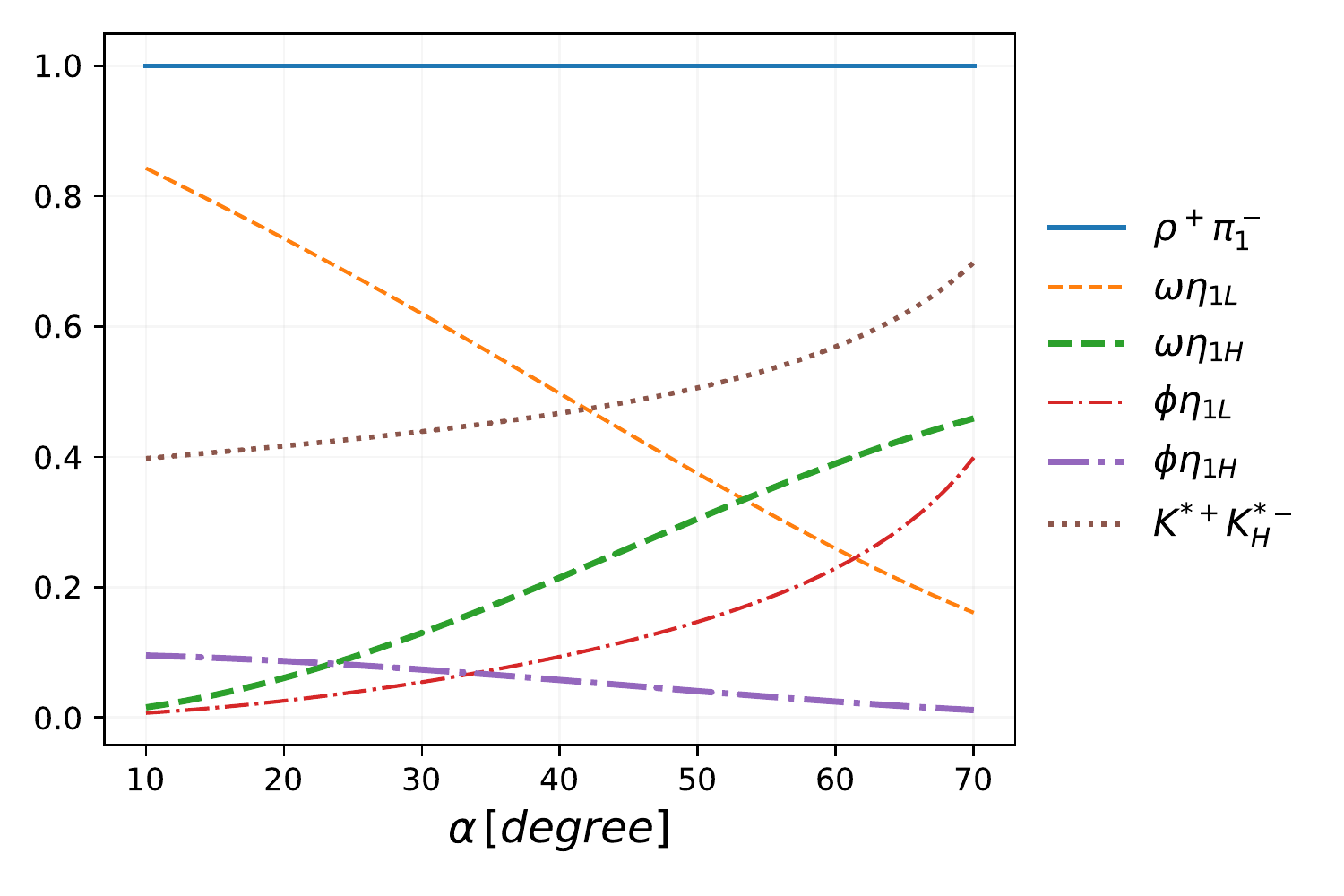}
\caption{Predicted branching ratio fractions for $BR(J/\psi\to VH)/BR(J/\psi\to \rho^+\pi_1^-)$ in terms of $\alpha$ in Scheme-I. }
\label{fig-ratio-VH}
\end{figure}

In Fig.~\ref{fig-ratio-VH}, we plot the branching ratio fractions in Scheme-I for each channels in respect with $J/\psi\to\rho^+\pi_1^-$ for which we set it as unity, and the other channels are normalized to the $\rho^+\pi_1^-$ channel with $R=0.93$ adopted. The favored range of the mixing angle is about $\alpha\in (17^\circ, 43^\circ)$. But a relatively broad range is plotted in  Fig.~\ref{fig-ratio-VH} as an illustration. The dependence of the mixing angle produces certain patterns which makes the combined study of the $VH$ channel useful for further confirmation of the $1^{-(+)}$ nonet. 

For the results of Scheme-II, the $\omega\eta_{1H}$ and $\phi\eta_{1H}$ thresholds are higher than the $J/\psi$ mass. A combined study can be done in other higher heavy quarkonium decays such as $\Upsilon\to VH$.

\section{Summary}

Inspired by the observation of the isoscalar $1^{-+}$ hybrid candidate $\eta_1(1855)$ in $J/\psi\to \gamma\eta_1(1855)\to\gamma\eta\eta'$, we investigate its SU(3) flavor partners using a parametrization method based on the flux tube model picture. We show that, although the present experimental information is still limited, it is possible to depict the $1^{-(+)}$ nonet of which the production and decays are consistent with the expectations of the flux tube model. We find that the observation of a single $\eta_1(1855)$ in the $\eta\eta'$ channel is informative and can impose quite strong constraint on the hybrid scenario. In particular, it suggests that the flavor octet and singlet mixing would be apparently deviated from the SU(3) ideal mixing which indicates the importance of the quark annihilation effects. In the flux tube model, it suggests that the transverse mode of the flux tube motions is important. 

We examine two schemes for the $1^{-(+)}$ hybrid nonet by assigning the observed $\eta_1(1855)$ to be either the high or low mass state with $I=0$. In both cases, we find that the requirement that one $I=0$ state should be highly suppressed in $J/\psi\to\gamma\eta\eta'$ will also impose a strong constraint on the hybrid $K^*$ mass. For the case that $\eta_1(1855)$ is the higher mass state, $K^*(1680)$ seems to be able to fill the nonet chart reasonably well. For $\eta_1(1855)$ being the lower mass state, a new state $K^*(1860)$ is predicted. We find that one of the main differences between these two solutions lies on the relative sign between the transverse and longitudinal mode of the gluonic motions in the decay of the $I=0$ hybrid states. It means that additional constraint from other processes are needed. As a strongly correlated process, we suggest a combined study of $J/\psi \to VH$ to clarify the difference between these two schemes. Notice that the $\rho\pi_1$ production rate is expected to be significant and specific pattern actually arises from the hybrid scenario. Thus, further evidences for $\eta_1(1855)$ and its partner in $J/\psi \to VH$ at BESIII would be crucial for finally establishing the $1^{-(+)}$ nonet.

\begin{acknowledgments}
The authors thank Xiao-Hai Liu and Alessandro Pilloni for useful comments and discussions.
This work is supported, in part, by the National Natural Science Foundation of China (Grant Nos. 11425525 and 11521505),  DFG and NSFC funds to the Sino-German CRC 110 ``Symmetries and the Emergence of Structure in QCD'' (NSFC Grant No. 12070131001, DFG Project-ID 196253076), Strategic Priority Research Program of Chinese Academy of Sciences (Grant No. XDB34030302), and National Key Basic Research Program of China under Contract No. 2020YFA0406300.

\end{acknowledgments}

\bibliographystyle{unsrt}

\end{document}